\documentclass[article,twocolumn]{revtex4}
\usepackage{amsmath,amssymb,graphicx,latexsym}

\newcommand{\be}{\begin{equation}}
\newcommand{\ee}{\end{equation}}
\newcommand{\bea}{\begin{eqnarray}}
\newcommand{\eea}{\end{eqnarray}}

\newcommand{\vev}[1]{\left<{#1}\right>}
\newcommand{\bra}[1]{\left|{#1}\right>}
\newcommand{\ket}[1]{\left<{#1}\right|}

\begin{document}
\title{Quantum dual-path interferometry scheme for axion dark matter searches}

\author{Qiaoli Yang$^{1}$\footnote{Contact author: qiaoliyang@jnu.edu.cn}}
\author{Yu Gao$^{2}$\footnote{Contact author: gaoyu@ihep.ac.cn}}
\author{Zhihui Peng$^{3}$\footnote{Contact author: zhihui.peng@hunnu.edu.cn}}

\affiliation{
$^{1}$College of Physics and Optoelectronic Engineering, Department of Physics, Jinan University, Guangzhou 510632, China\\
$^{2}$Key Laboratory of Particle Astrophysics, Institute of High Energy Physics,\\
Chinese Academy of Sciences, Beijing 100049, China\\
$^{3}$Key Laboratory of Low-Dimensional Quantum Structures and Quantum Control of Ministry of Education, Key Laboratory for Matter Microstructure and Function of Hunan Province, Department of Physics and Synergetic Innovation Center for Quantum Effects and Applications, Hunan Normal University, Changsha 410081,
China
}


\begin{abstract}
Exploring the mysterious dark matter is a key quest in modern physics. Currently, detecting axions, a hypothetical particle proposed as a primary component of dark matter, remains a significant challenge due to their weakly interacting nature. Here we show at quantum level that in a cavity permeated by a magnetic field, the single axion-photon conversion rate is enhanced by the cavity quality factor and is quantitatively larger than the classical result by $\pi/2$. The axion cavity can be considered a quantum device emitting single photons with temporal separations. This differs from the classical picture and reveals a possibility for the axion cavity experiment to handle the signal sensitivity at the quantum level, e.g., a dual path quantum interferometry with cross-power and second-order correlation measurements. This scheme would greatly reduce the signal scanning time and improve the sensitivity of the axion-photon coupling, potentially leading to the direct observation of axions.
\end{abstract}

\date{\today}

\maketitle
\section{Introduction}
The existence of cold dark matter is widely accepted~\cite{Ade:2015xua}. One promising candidate is the quantum chromodynamics (QCD) axion~\cite{Peccei:1977hh,Peccei:1977ur, Weinberg:1977ma,Wilczek:1977pj,  Kim:1979if,Shifman:1979if,Zhitnitsky:1980tq,Dine:1981rt,Preskill:1982cy,Abbott:1982af,Dine:1982ah,Sikivie:1982qv,Wilczek:1991jgb} which is a natural extension of the standard model of particle physics. The standard model predicts that the amount of Charge-Parity (CP) violation of the strong interaction should be order one. However, experiments measuring the neutron electric dipole moment~\cite{Crewther:1979pi} suggested an essentially vanishing CP violation~\cite{Baker:2006ts}, which gives rise to a puzzle now known as the strong-CP problem~\cite{Peccei:2006as,Kim:2008hd}. The problem could be solved by adding the Peccei-Quinn (PQ) symmetry~\cite{Peccei:1977hh,Peccei:1977ur}, which is spontaneously broken by PQ preserving potential and is explicitly broken by the QCD instanton effect~\cite{Callan:1977gz, Vafa:1984xg}. The symmetry breaking results in a pseudo Nambu-Goldstone boson, the QCD axion \cite{Wilczek:1991jgb}. Currently, there are two particular QCD axion benchmark models, \cite{Kim:1979if, Shifman:1979if} and \cite{Zhitnitsky:1980tq, Dine:1981rt}. Cosmic axions can be produced abundantly during the QCD phase transition in the early universe, and subsequently their relics could compose the cold dark matter observed today~\cite{Preskill:1982cy,Abbott:1982af,Dine:1982ah,Sikivie:1982qv}. Depending on whether the PQ symmetry occurred before or after cosmological inflation, the natural mass window of the QCD axions being the majority of dark matter is approximately ${\cal O}(10^{-5}- 10^{-3})$eV~\cite{Sikivie:2006ni} or $<{\cal O}(10^{-7})$ eV~\cite{Hertzberg:2008wr}. However, other production mechanisms are possible, which will lead to a wider mass window. QCD axions couple to the Standard Model particles weakly; in particular, it couples to two photons with a coupling ${\cal O}(10^{-17} - 10^{-12})~$GeV$^{-1}$. Currently, many proposed and ongoing experiments such as~\cite{DePanfilis:1987dk,Hagmann:1990tj,Kahn:2016aff,Caldwell:2016dcw,HAYSTAC:2018rwy,Marsh:2018dlj,Schutte-Engel:2021bqm,ADMX:2021nhd} are actively searching the dark matter axions. In condensed-matter systems, quasiparticle "axions" which lead to many exciting phenomena \cite{Wilczek:1987mv,Essin:2008rq,Li:2009tca,Xiao:2017mol,PhysRevLett.105.190404,Nenno:2020ujq} have already been observed. It is tempting to believe that nature may repeat itself at many different levels.

Axion haloscope experiments are generally based on resonant cavity design~\cite{Sikivie:1983ip}, in which the cosmic axions resonantly convert into a microwave signal in a high quality factor, $Q$, cavity permeated by a magnetic field. Although the axion-photon coupling is small, the conversion rate is enhanced by the coherence of the axion field, the magnitude of the magnetic field, and the high $Q$ of the cavity. Modern cryogenic technology can sustain ${\cal O}(20)$ mK or lower temperature, which results in a subunity thermal photon occupation number $\bar n\sim 10^{-5}$ at the resonant frequency (typically at several GHz) in the cavity. Therefore, it is useful to consider the quantum transition picture in the cavity.

In this paper, we propose to use a 50/50 beam splitter followed by linear amplifiers or single photon detectors to construct a dual-path measurement scheme that realizes Hanbury Brown and Twiss (HBT) interferometry~\cite{Brown:1956zza} in quantum optics. Linear detectors (phase-preserving amplifiers) are widely used in axion haloscope experiments. Current commercial cryogenic high electron mobility transistors (HEMT) already offer a flat gain over a frequency range of $10\,$GHz with the addition of random noise at 10-20 photons, and amplifiers such as the Josephson Parametric Amplifier (JPA)~\cite{JPA2011} can approach the standard quantum limit (SQL). At very low signal/thermal photon occupation numbers, the major sensitivity hurdles are due to quantum fluctuations and added noise in the detection channel. The traditional method of signal identification is to accumulate sufficient statistics at each given frequency point, so it takes a long integration time due to the unfavorable signal-to-noise ratio. The HBT interferometer scheme deployed in microwave quantum optics, however, has already been demonstrated to achieve {\it high} noise reductions~\cite{Menzel2010,Bozyigit2010,Peng2016,Zhou2020} because the uncorrelated noises in the dual receiver chains cancel and the correlations of field quadratures can be measured. With this scheme, the effective correlated noise level can be significantly lower than the noise level in a single amplification channel, as the interferometer setup is not sensitive to the inevitable microwave signal insertion loss in the channels between the cavity and the amplifiers.

\section{Methods}
\subsection{Axion cavity at the quantum level}
The viable QCD axion models predict extremely weak couplings and light masses. Fortunately, the dark matter axion production mechanism \cite{Preskill:1982cy,Abbott:1982af,Dine:1982ah,Sikivie:1982qv} suggests a preferred axion mass window around neV which could be the searching target, even though the uncertainty still is several orders of magnitude. Many experimental schemes aimed at this window have been proposed. In these proposals, the axion cavity haloscope is particularly sensitive and can reach the QCD axion parameter space within current technology. The original calculation was given \cite{Sikivie:1983ip,Sikivie:1985yu} in a picture in which the cavity captures the photons produced and enhances the conversion process through resonance when the cavity modes match the energy of the axion particle. However, it was uncertain whether the transition rate can be enhanced during a single axion-photon conversion, which is quantum in nature. Some calculations, e.g., Ref.~\cite{Beutter:2018xfx}, applied the Feynman diagram method, which requires the existence of asymptotic outgoing states and thus cannot derive the resonant cavity's enhancement and form factor etc.

The axion field $a$ couples to the photons by
\be
{\cal L}_{a\gamma\gamma}=-g_{a\gamma\gamma}a\vec E\cdot \vec B~~,
\ee
where the axion-photon coupling $g_{a\gamma\gamma}$ is defined as $g_{a\gamma\gamma}=c_{\gamma}\alpha / (\pi f_a)$, $f_a$ is the PQ scale factor, $\alpha=1/137$ is the fine-structure constant and $c_{\gamma}$ is a model-dependent factor order of one.

The axion dark matter halo profile is crucial to axion searches. There are several important parameters for the current discussion: the dark matter halo local density $\rho_a=\rho_{CDM}\approx 0.45$GeV/cm$^3$; the axion dark matter coherent time, which is relatively long due to the small axion mass $m_a$, the slow velocity $v_a\sim 10^{-3}$c and the small velocity distribution $\delta v_a\leq v_a$.

Due to the small mass and slow velocity, the axion dark matter has a long de-Broglie wavelength $\lambda_a=2\pi/(m_av)\sim {\cal O}(1 - 100)$ meters, which is much larger than the size of a respective resonant cavity, so the axion field is homogeneous in a haloscope cavity. In addition, the axion coherence time $\tau_a=\lambda_a/\delta v_a> 2\pi/(m_av_a^2)$ is significantly longer than the photon existence time in the cavity: $\tau_c=2\pi Q/\omega_a\approx 2\pi Q/m_a$, where $Q\sim 10^5$ for current state-of-the-art technology; thus, the axion field can be regarded as monochromatic during a cavity model transition (note that in some axion halo models, the coherence time
could be even longer if $\delta v_a\ll v_a$ \cite{Sikivie:2001fg, Armendariz-Picon:2013jej}). Thus, we can write the axion field as
\be
a\approx a_0{\rm cos}(\omega_at)={\sqrt{2\rho_a}\over m_a}{\rm cos}(\omega_a t)~.
\ee
In the experimental literature, the axion quality factor $Q_a={1/ v_a^2}\sim 10^6$ is often used to describe the coherence of axions.

In a cavity, the electric field operator $\vec E$ can be expanded as
\bea
\vec E=i\sum\sqrt{\omega_k\over 2}[a_k\vec U_k(\vec r)e^{-i\omega_kt}-a^\dag_k \vec U^*_k(\vec r)e^{i\omega_kt}],
\eea
where $a_k$, and $a^\dag_k$ are the annihilation and creation operators for the photon Fock state respectively, and $\vec U_k(\vec r)$ are the cavity modes satisfying the wave equation $(\nabla^2+\omega_k^2)\vec U_k(\vec r)=0$ with the cavity-wall boundary conditions. Assuming that the permitted magnetic field is along the $\hat z$ direction, $\vec B=\hat z B_0$, the interaction Hamiltonian can be written as
\bea
H_I&=&-\int d^3x {\cal L}_{a\gamma\gamma} \nonumber \\ &=&\left(g_{a\gamma\gamma}{\sqrt{2\rho_{a}}\over m_a}B_0\int d^3x\hat z\cdot \vec E\right){\rm cos}(\omega_at)~~.
\eea
Up to the first order, the photon $\bra 0  \to \bra1$ transition probability is
\bea
P&\approx&\left|\ket 1\int_0^tdtH_I\bra 0 \right|^2\nonumber\\&\approx&g^2_{a\gamma\gamma}{\rho_{a}\over m_a^2}B^2_0\sum_k\omega_k|\int d^3x\hat z\cdot\vec U^*_k|^2\nonumber\\&\times&{{\rm sin}^2[(\omega_k-\omega_a)t/2]\over4 [(\omega_k-\omega_a)/2]^2}~~.
\eea
When $t$ is large compared to $|1/(\omega_k-\omega_a)|$, we can use the approximation ${\rm sin}^2(\Delta \omega t/2)/[\Delta\omega /2]^2\approx 2\pi t\delta(\Delta\omega)$ \cite{Fitzpatrick2015}. The transition rate is $R={dP/dt}$. Let us define the form factor
\be
C_k={\left|\int d^3x \hat z\cdot\vec U_k\right|^2\over V\int d^3x|\vec U_k|^2}~,
\ee
where $V$ is the volume of the cavity, and the factor $\int d^3x|\vec U_k|^2$ properly normalizes the photon field operator.
Then we have
\bea
R&\approx&{\pi\over 2}g_{a\gamma\gamma}^2{\rho_{a}\over m_a^2}B_0^2 V\sum_k C_k\omega_k\delta(\omega_k-\omega_a)~~.
\eea
Since $\sum C_k\omega_k\delta(\omega_k-\omega_a)\approx\int C_k d\omega (\omega/d\omega)\delta(\omega-\omega_a)\approx QC_{\omega_a}$, we have the transition rate
\be
R\approx {\pi\over 2}g_{a\gamma\gamma}^2{\rho_{a}\over m_a^2}B_0^2C_{\omega_a} VQ~~.
\ee
This shows that the single axion-photon transition power $P_{sig}=\omega_aR\approx m_a R$ is {\it enhanced} by the cavity's quality factor $Q$, which is in good agreement with the calculation in the classical picture save for a quantum $\pi/2$ forefactor.
For a typical haloscope setup, the cavity volume is approximately one liter, with $Q\sim 10^5$ and $B\sim 10$ T; then, one finds that the $a\rightarrow \gamma$ conversion rate is approximately ${\cal O}(1)$ per second for the QCD axions. The signal temporal separations far exceed the resolution of linear detectors.

During laboratory measurements, the axions are free-streaming dark matter, so that \cite{Foster:2017hbq} "the individual axion particle number is conserved and the phase coherence of the full axion field constructed from the sum of these particles is dominated by the common mass they share and to a lesser extent by velocity corrections which are drawn from a common dark matter (DM) velocity distribution. Beyond this the fields are entirely uncorrelated." Thus, the axion DM field is different from the classical coherent field, such as the electric field. When the axion DM is converted into photons, the axions still behave as particles with a fixed mass and particle number. Due to energy conservation, each time, only one axion can be converted into one photon with energy mostly dominated by axion mass (the process of double-conversion is highly suppressed in comparison, as we will explain later).

The thermal photon occupation number in the cavity is
\bea
n(\omega_a, T)={1\over e^{\omega_a/k_BT}-1} ~~.
\eea
For axion mass $m_a\ge 10^{-5}$eV and a cavity operating at approximately $T\approx 20$ mK, the thermal photon state has a low occupation number {$n\ll 1$}; thus, the cavity is almost always in the vacuum state. It is convenient to define the cavity coupling parameter $\beta=Q_0/Q_c$, where $Q_0$ is the quality factor due to the photon losses by cavity-wall absorption and $Q_c$ is due to the photons leaving the cavity. The combined quality factor is then $Q^{-1}=Q_0^{-1}+Q_c^{-1}$, and the photon emission rate is $R\cdot\beta/(1+\beta)$. Ideally, $\beta\sim 1$ is the optimal working point, but its exact value depends on the cavity design.

The single-photon emitting nature of the cavity can be observed. First, the probability of converting multiple axions into multiple photons via higher-order processes is very small. Generally, for example, there are two ways to achieve two-photons conversion in a single process: (1) $a a\rightarrow \gamma \gamma$ annihilation with the exchange of a virtual photon. This process is suppressed by the $g_{a\gamma\gamma}^4$ factor in comparison to the $g_{a\gamma\gamma}^2$ factor in single-photon conversion. In addition, the annihilation process does not connect to the external magnetic field, so it does not receive an enhancement from the external magnetic field. (2) Higher-order operators that convert two axions into two photons under an external magnetic field. The lowest-order gauge-invariant of such operators will take the form $a^2\left[(FF)^2+(F\tilde{F})^2+...\right]$, where the quartic field terms can assume different coefficients. These quantum operators are dimension-10; compared to axion-axion annihilation, they suffer even more severe suppression. Thus, such a contribution can be safely ignored for axion-photon conversions in the cavity.

Second, although there is a chance that two or more single axion-photon conversions occur before the converted photons exit the cavity, these probabilities are extremely small. After a photon appears, the duration of its existence in the cavity is $~2\pi Q/m_a$. During this period, the probability of another axion photon conversion is $R\cdot ~2\pi Q/m_a\ll 1$.

Certainly, the signal rate drop to the single-photon level is not the same as the rate at which photons are deterministically released. In principle, a photon state $\alpha \bra 0+\beta \bra 1$ (where $\alpha$ and $\beta$ are complex numbers and hold the normalized condition $|\alpha|^2+|\beta|^2=1$) could exist, but it would become almost "decoherent". The B-field induced photon-axion transition is extremely weak, so once a photon is generated inside the cavity, it cannot transit back to a $\bra 0$ state before the readout measurement occurs, as the latter is much faster and effectively breaks the "coherence". 

Furthermore, one could consider a similar scenario: the Purcell effect, i.e. \cite{Qian:2021}. The ability of the Purcell effect to control the spontaneous emission rate through cavity modes is one of the key principles of realizing a single-photon source. Axion dark matter with a static magnetic field B can be regarded as a large number of independent, weakly coupled (to cavity) two-level systems since the transition rate of axion(s) to multiple photons is much lower than the one-photon transition. The photon-cavity interaction enhances the axion-photon transition rate as does the Purcell effect. Therefore, it is possible to consider an alternative particle picture in which axion dark matter particles can be regarded as radioactive isotopes that spontaneously decay into photons. The existence of a cavity enhances the decay rate because the phase space of decayed photons is constrained. The wave and the particle pictures are complimentary in this case.

Combining these considerations, the axion-converted signal is very close to an ideal single-photon pulse. In accordance with the current experimental parameters, the axion cavity can be regarded as a single photon emitter with a slow rate of $R\lesssim 10$ Hz.

\subsection{Dual-path interferometry}
Since the axion cavity is a single-photon emitter at low pulse rates, the major experimental bottleneck lies in the detection sensitivity of weak microwave signals at the single-photon level. To date, practical microwave single-photon detectors have yet to be realized~\cite{PhysRevX.10.021038,Blais:2020wjs,PhysRevLett.126.141302}. The main challenge is that the microwave photon energy is typically 4-5 orders of magnitude lower than that of an optical photon, and it is difficult for detectors to distinguish single-microwave photons from large noise backgrounds~\cite{PhysRevX.10.021038}. In the future, the integration of materials science, advanced nanofabrication techniques, and quantum technologies will be essential for realizing practical microwave single-photon detectors. Currently, a typical linear detector scheme for an axion cavity search consists of an amplification chain, which is a cryogenic HEMT amplifier as a preamplifier placed at the $4\,$K temperature stage in a dilution refrigerator, and subsequent room-temperature amplifiers. The amplified microwave signal at GHz was downconverted to a few tens of MHz and then sampled by digitizers at room temperature. The typical noise temperature of a cryogenic HEMT is approximately several Kelvin. Considering the loss in the channel from the microwave signal emitted from the cavity at the $20\,$mK stage to the cryogenic HEMT placed at the $4\,$K stage in the refrigerator, the effective noise temperature $T_{eff}$ in the detection channel could be higher than $10\,$K. With the development of superconducting quantum information technology, recent axion cavity experiments, such as ADMX-Sidecar~\cite{Bartram:2021ysp}, QUAX-$a\gamma$~\cite{Alesini:2020vny}, etc., have applied TWPAs or JPAs, respectively, as near quantum-limit preamplifiers to enhance detection sensitivity. However, there is possible deterioration of the superconducting JPA in a strong magnetic field environment, and there is approximately $\sim-3\,$dB insertion loss from coaxial cables, switchers and circulators placed between the cavity and the quantum-limited amplifiers $T_{eff}$ of the detection channel, which is approximately four times higher than the SQL~\cite{Bartram:2021ysp}.

\begin{figure}
\textbf{Scheme of the quantum dual-path setup for a resonant cavity dark matter axion search.}\par\medskip
\includegraphics[width=0.5\textwidth]{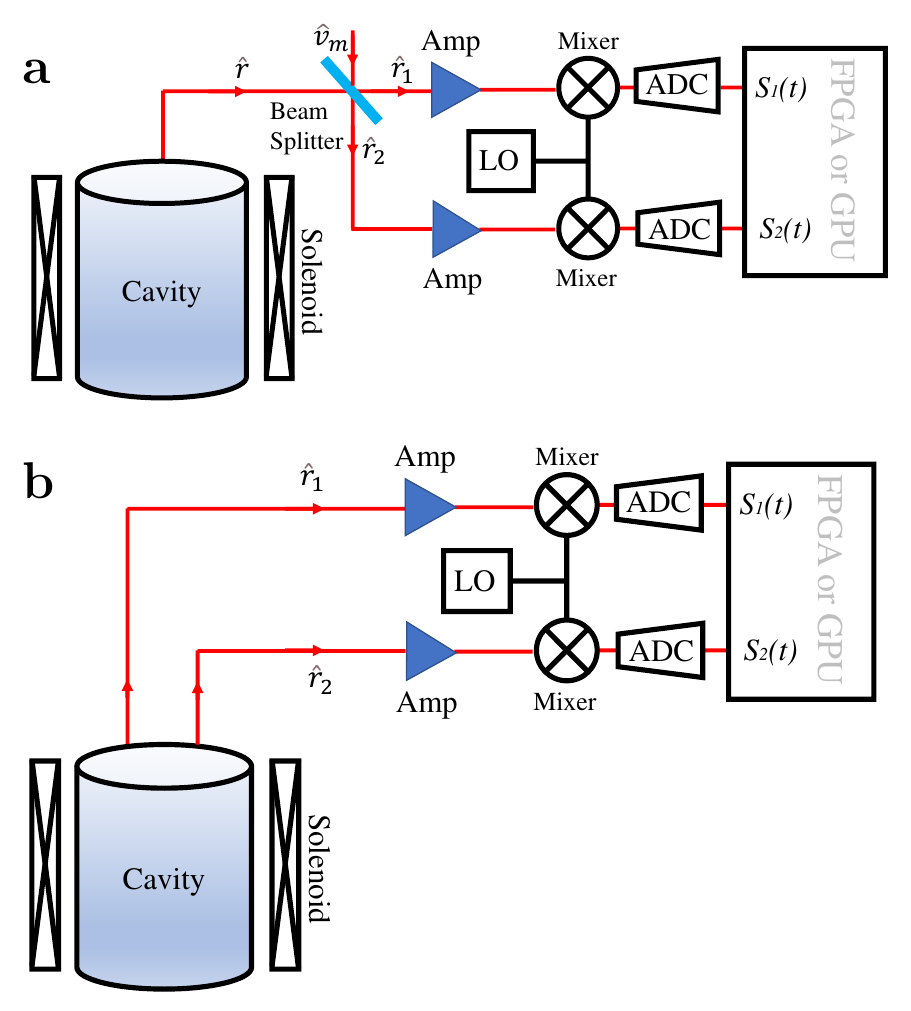}
\caption{(a) Single-sided cavity scheme. There is only one output port (outgoing red line) of the cavity detector. Axions convert to microwave single photons that escape from the cavity as a propagation quantum field with an annihilation operator $\hat{r}$ and subsequently pass through a 50/50 beam splitter along with a vacuum field or weak thermal field $\hat{v}_{m}$ near the cavity at the $20\,$mK stage in a refrigerator. The output fields $\hat{r}_{1}$ and $\hat{r}_2$ are then amplified by two nominally identical amplification chains (denoted as AMP, blue triangle) and downconverted into intermediate (e.g., $25\,$MHz) frequency $\omega_{if}$ with a mixer by the local oscillator (LO) at the same frequency as the cavity's resonance frequency. The field quadratures are recorded by analog-to-digital converters (ADCs), and the complex envelopes $S_i$ are then extracted and calculated by FPGA or GPU electronics in real time. The single-sided cavity scheme can overcome the inevitable insertion loss between the beam splitter and the AMP. (b) Two-sided cavity scheme. If there are two identical output ports of the cavity detector, the converted microwave single photons escape from the two output ports with equal probability (outgoing red lines), and a beam splitter is not necessary. The correlations between the two cavity outputs behave similarly to the outputs of the beam splitter in (a). However, it is expected that the SNR in this scheme is even higher than that in the scheme described in (a). It can overcome the inevitable insertion loss between the cavity and the AMP.}
\label{SchemeFig}
\end{figure}

Thus, the effective temperature $T_{eff}$ on readout is often much higher than the physical temperature in the cavity. The signal-to-noise ratio is
\be
{\rm SNR}={P_{sig}\over k_B T_{eff}}{\sqrt{t\over b}}~~,
\label{eq:snr}
\ee
where $b$ is the detection bandwidth, $P_{sig}$ is the signal power, $k_B$ is the Boltzmann constant and $t$ is the integration time. The sensitivity on $g^{2}_{a\gamma\gamma}$ is inversely proportional to $T_{eff}$ or grows over the square root of the integrated time.

Here, we briefly interpret the quantum noise with detector observables,
\be
\hat I_r\cos(\omega_at)+\hat Q_r\sin(\omega_at)={\rm Re}\left((\hat I_r+i\hat Q_r)e^{i\omega_at}\right)
\ee
where we define the in-phase $\hat I_r$ and quadrature $\hat Q_r$ component operators with $\hat I_r=(\hat{r}^{\dag}+\hat{r})/2,~\hat Q_r=-i(\hat{r}^{\dag}-\hat{r})/2$ using the creation and annihilation operators $\hat{r}^{+}$ and $\hat{r}$. These operators satisfy commutation relations
\be
[\hat I_r,~\hat Q_r]=i/2~~{\rm and}~~ [\hat{r}^{+},\hat{r}]=1~~.
\ee
Assuming the linear detector enhances the field by a gain of $G$, we have
\be
{\hbar\omega_a\over 2}+{(G^2-1)\hbar\omega_a\over 2 G^2}\approx \hbar\omega_a,
\ee
and for a large $G$, this is often called the SQL~\cite{Caves1982}. To protect the cavity at $20\,$mK from thermal noise leakage from the second-stage amplifier (e.g., cryogenic HEMT at 4 K), a preamplifier gain over $20\,$dB is needed. To the best of our knowledge, it is still difficult to achieve 20 dB gain over a 1 GHz bandwidth for a JPA or a traveling-wave parametric amplifier (TWPA)~\cite{Macklin2015}, with the noise level approaching the SQL simultaneously.

For signals with quantum origin, a dual-path measurement can be a powerful tool because it measures the statistical correlation between quadratures in the two channels in addition to the photon number $n=\vev{a^\dag a}$ in each channel.
Recent quantum optics developments enable dual-path detection in the microwave-frequency domain, and it can be performed with phase-preserving linear amplifiers as well~\cite{Menzel2010,Bozyigit2010,Peng2016,Zhou2020,Eichler2015}.
Linear amplifiers are well known for their high signal gain for weak electromagnetic field signals but at the cost of extra noise.
With a 50/50 beam splitter, as shown in Fig.~\ref{SchemeFig}, dual linear amplifier chains can be constructed to measure the signal correlation information. It has been shown~\cite{Menzel2010,Bozyigit2010,Zhou2020} that with a recording of the full time traces of the signals in both channels instead of the time-averaged value in a single channel, the signal's cross power in two channels can be extracted and significantly improves the SNR.

The dual-path scheme in Fig.~\ref{SchemeFig}(a) forms a Hanbury Brown-Twiss interferometer setup. The axion cavity acts as a quantum emitter to inject a microwave single-photon field $\hat r$, which passes through a microwave-frequency beam splitter. The beam splitter gives rise to two output fields in channels $1$ and $2$,
\be
\hat r_1=(\hat r+\hat\nu_m)/\sqrt2~~{\rm and}~~\hat r_2=(\hat r-\hat\nu_m)/\sqrt2~~.
\ee
$\hat\nu_m$ is the added noise field, for instance, $\hat{\nu}_m$ as the vacuum state or a weak thermal state. Then, by measuring $\hat I_1=(\hat r_1+\hat r_1^+)/2$ and $\hat Q_2=-i(\hat r_2-\hat r_2^+)/2$, a complex observable can be constructed as
\be
\hat S_{a}(t)=I_1(t)+iQ_2(t)=\hat r+\hat \nu_m^{+}~~.
\ee In fact, $\hat S_{a}(t)$ behaves classically and resembles a complex number because $\hat S_{a}^+=\hat S_{a}^*$ and $[\hat S_{a}^+,\hat S_{a}]=0$ under the effect of the added noise $\nu_m$~\cite{Silva2010}. After amplification and mixing, the two classical complex envelopes $S_{i}(t)$, $i=1,2$, at the digitizers can be written as
\be
S_{i}(t)=G_i\hat r(t)+ \sqrt{G_i^2-1}h_i^+(t)+\nu^{+}_{m,i}(t)~~,
\ee
where we let $h_i$ denote the added thermal noise in each linear amplifier chain and $\nu_{m,i}(t)$ denote the vacuum noise. Both quadratures of $\hat S_i$ can be measured simultaneously by digitizers due to the commutator $[\hat S_i, \hat S_i^+]=0$. A powerful Field Programmable Gate Array (FPGA)~\cite{Menzel2010,Bozyigit2010} or Graphics Processing Unit (GPU)~\cite{Zhou2020} can analyze the correlation of field quadratures in real time. The correlations of the output field quadratures are constructed as
\be
\vev {(S_i^{*})^mS_j^{n}}=\vev {(S_i^{+})^mS_j^{n}}=\vev {(\hat r^{+})^m\hat r^{n}}~~,
\ee
with integer-number power indexes $m$ and $n$.
\begin{figure}
\textbf{Comparison between simulated cross power and single-channel power.}\par\medskip
\includegraphics[width=0.45\textwidth]{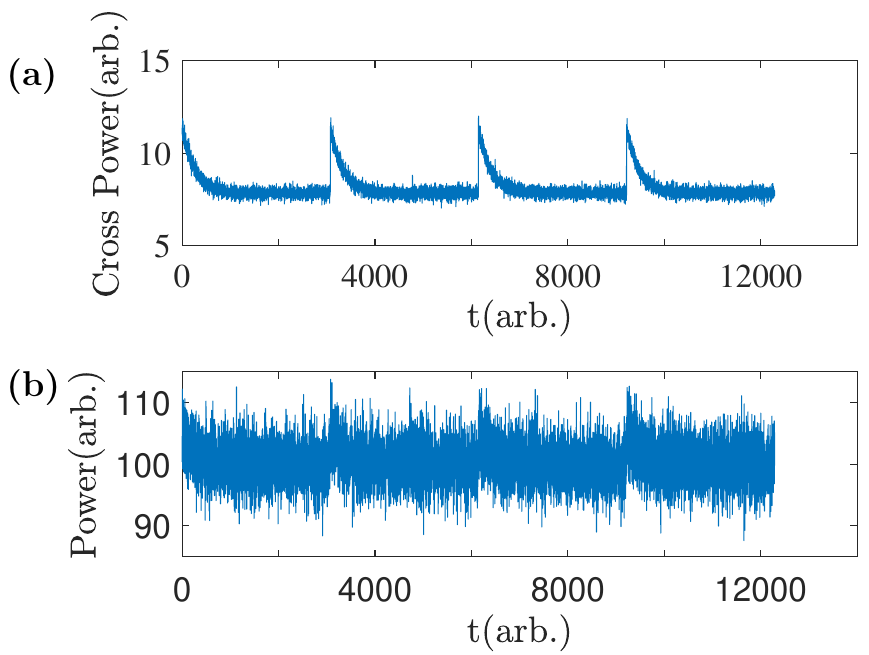}
\caption{(a) The signal-to-noise ratio is much higher when using the cross-power setup, thus revealing the injected signal. (b) A periodical signal is injected with white noise. As described in Eq.~\ref{eq:snr}, the scanning time is proportional to the square of the SNR. We expect the signal scanning time to be greatly reduced with the dual-path scheme.}
\label{TimeTraceFig}
\end{figure}

\section{Results}
\subsection{Sensitivity enhancement}
It is now straightforward to show that the signal can be picked up by the instantaneous power function $\vev{S_i^*(t)S_j(t)}$, where the {\it power} is
\be
\vev {S_i^*(t)S_i(t)}=G_i^2\left(\vev{\hat r^+(t)\hat r(t)}+N_i+0.5\right)~~,
\ee
and the {\it cross-power} is
\be
\vev {S_1^*(t)S_2(t)}=G_1G_2\left(\vev{\hat r^+(t)\hat r(t)}+N_{12}+0.5\right)~~.
\ee
Here, the component $P_{\rm cavity}=\vev{\hat r^+(t)\hat r(t)}$ identifies the power from the cavity, $N_i=(G_i^2-1)\vev{h_i^+h_i}/G_i^2$ is the power of noise added in a single channel, and $N_{12}=\sqrt{(G_1^2-1)(G_2^2-1)}\vev{ h_1h_2^+}/G_1G_2$ is the power of correlated noise between channels 1 and 2. $\vev{ h_1h_2^+}\sim0$ because the two noise modes in the two detection channels are commutable and mostly uncorrelated. That is,
\be
N_{12}+0.5\ll N_{i}+0.5~~,
\ee
and the magnitude of dual-path noise is significantly lower than the magnitude of single-path noise, as illustrated in Fig~\ref{TimeTraceFig}. Notably, $N_{12}+0.5\geq0.5$ and $N_i+0.5\geq1$ are related to the noise set by SQL. The SNR in the dual-path scheme is enhanced two times compared to that in the single-path scheme only in the ideal case. We emphasize that it is crucial to upgrade from the traditional single-path setup to {\it dual-path} to cancel the main part of uncorrelated noises from different detection channels in real experiments. In practice, $N_i$ is almost larger than 4 because of the possible deterioration of the superconducting JPA in a strong magnetic field environment and the microwave signal insertion loss between the cavity and the detector in the current single-path experiment with JPAs working in the GHz frequency range~\cite{Bartram:2021ysp}. If the search frequency range is lower than $1\,$GHz, $N_i$ is even larger than 10 in the real experiment~\cite{ADMX:2021nhd}. However, the scale of the remaining correlated noise $N_{12}$ depends on the experimental devices and can be suppressed to much smaller than 0.5, and it can be directly measured by the cross power without an axion cavity. This can be explained by the thermal leakage from the cryogenic HEMT at $4\,$K inducing correlated noise between the two channels of the beam splitter. Actually, it has been demonstrated that the correlated noise $N_{12}$ can be reduced and the SNR of the dual-path scheme can be enhanced for the cross-power measurement if there are quantum isolators or JPAs between the beam splitter and HEMT at $4\,$K \cite{Eichler2015}.

\begin{figure}
\textbf{Simulation of the second-order correlation function for a single-photon pulse train with white noise.}\par\medskip
\includegraphics[width=0.45\textwidth]{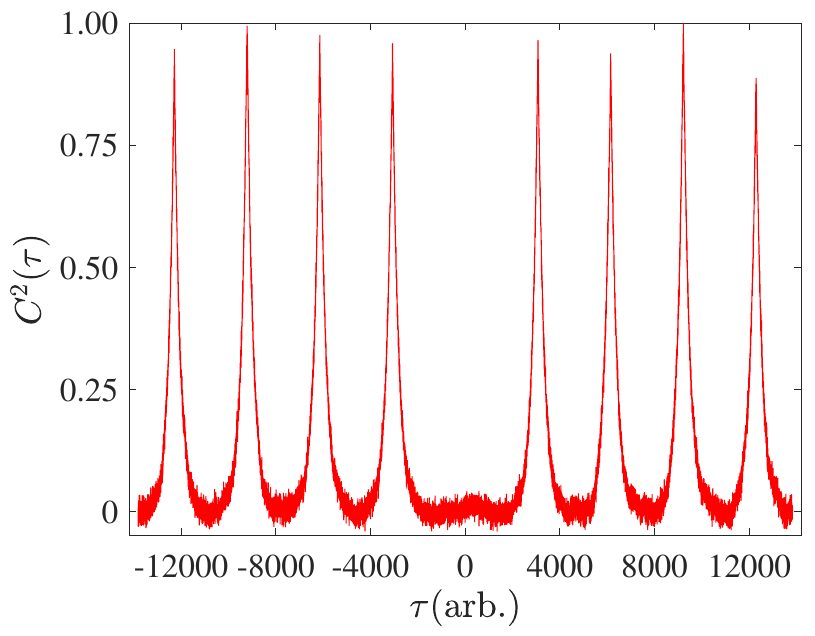}
\caption{If photon antibunching is observed in the experiment as $C^2(0)<C^2(\tau)$, the converted periodic signal is nonclassical.}
\label{G2Fig}
\end{figure}

The preamplifiers or detectors can be placed at a distance from the cavity output port to help shield the strong magnetic field. Insertion loss is inevitable because there are a series of microwave coaxial cables, switchers and circulators between the cavity port and the preamplifiers/detectors. In realistic experiments, for instance, the effective noise temperature in the detection channel is $\sim 925\,$mK at $4.798\,$GHz even when using JPA with an insertion loss $\sim-3\,$dB~\cite{Bartram:2021ysp}. However, the effective noise temperature should be $\sim 240\,$mK if it is only limited by the SQL. Therefore, there is an obvious advantage to using a dual-path interferometer scheme that can approach the effective noise temperature with $\sim 120\,$mK, which comes from quantum vacuum fluctuations, especially to use the two-sided cavity scheme described in Fig.1(b) to detect converted microwave single-photon signals. In the two-sided cavity dual-path scheme, there are two identical output ports, and the converted microwave single photons escape from them with equal probability. The correlations between the two cavity outputs behave similarly to the outputs of the beam splitter in Fig.~\ref{SchemeFig}(a). The dual-path interferometer is not sensitive to loss or thermal noise in the single channel, and the SNR could be enhanced by one order of magnitude compared with the single-path amplification scheme based on JPA.

Recent dual-path setups~\cite{Menzel2010,Bozyigit2010,Peng2016,Zhou2020} showed that by using cross-power, the effective noise temperature could be reduced compared to that of a single channel. Fig.~\ref{TimeTraceFig} illustrates the noise level reduction with the simulation of cross-power between two channels and the power in one channel averaged over the same time. It has been demonstrated that in a dual-path experiment with a cryogenic HEMT as preamplifiers, the correlated noise temperature is approximately $80\,$mK and the effective total noise temperature is approximately $180\,$mK, corresponding to half microwave photon noise, for the signal at $7.2506\,$GHz in the dual-path setup, which is much smaller than the characteristic $\sim10$ K noise in each detection chain~\cite{Bozyigit2010}.

By Eq.~\ref{eq:snr}, the SNR with a cross-power measurement is much higher than the power calculation by replacing the SQL with a reduced effective noise temperature. This allows faster signal-significance accumulation and reduces the amount of required time-exposure at each frequency point in future cavity axion dark matter search, esp. at relatively high frequencies. Actually, the assumption of equal gain in both amplification channels and a balanced microwave beam splitter is not difficult to achieve in experiments~\cite{Menzel2010,Bozyigit2010,Peng2016,Zhou2020}, and even an unbalanced detection channel does not hinder the cancellation of correlated noise.

Notice that when closing the quantum limit, using units of noise temperature to compare improvements between different schemes could lead to a large underestimate. For example, a one-eighth decrease in the noise temperature from 925mK to 120mK actually results in a 20 times improvement in units of photon:
\bea
\big({\rm{Exp}}[{{\hbar\omega}\over {k_b120\rm{mK}}}]-1\big)/\big({\rm{Exp}}[{{\hbar\omega}\over {k_b925\rm{mK}}}]-1\big)=20.
\eea
Researchers in particle physics may be more familiar with noise temperature, and the system noise temperature is directly related to the detection SNR. Therefore, we continue to quote the noise temperatures here.

\subsection{The second-order correlation function measurement}
For power correlation statistics, one may consider
\begin{eqnarray}
      C^2(t,t+\tau)&=&\vev {\hat r^{+}(t)\hat r^{+}(t+\tau)\hat r(t+\tau)\hat r(t)} \\\nonumber
            &=& 4\vev {S_1^{+}(t)S_1^{+}(t+\tau)S_2(t+\tau)S_2(t)}.
\end{eqnarray}
As widely exploited in quantum optics, single-photon signals demonstrate two important features: sub-Poissonian statistics $C^2(\tau)\leq 1$ and photon anti-bunching $C^2(0)<C^2(\tau)$~\cite{Bozyigit2010,Peng2016,Zhou2020,Eichler2015}.
When a signal candidate emerges during a scan, the second-order function measurement, as simulated in ~Fig.~\ref{G2Fig}, allows the measured $C^2(0)$ value to verify the nature of the signal: whether it is converted as a single photon from dark matter axion $(C^2(0)\sim 0)$ or simply from thermal noise $(C^2(0)\sim 1)$. $(C^2(0)\sim 0)$ arises from the fact that a single photon can only arrive at one path at one time, and it is a well-known technique in quantum optics to veto thermal noise~\cite{Menzel2010,Bozyigit2010,Peng2016,Zhou2020}. Thus, our dual-path scheme can provide an additional test of the signal's nature when the cavity's predicted occupation numbers are small, in which case most axion-converted signals exit the cavity as single photons.

We must emphasize that it is easy to combine the quantum dual-path scheme with other techniques, such as JPAs, microwave squeezing-state, and single-photon detectors. Even with JPAs, the single-path noise temperature is approximately four times higher than that limited by the SQL~\cite{Bartram:2021ysp}. However, with the dual-path measurement setup, noise due to the SQL in a single path could be avoided. With the squeezing state, a fractional breaching of SQL has been achieved~\cite{Backes2021}, while for cross-power, at least an order of magnitude difference between correlated and uncorrelated noise levels is typically expected~\cite{Menzel2010,Bozyigit2010,Peng2016,Zhou2020}. In addition, both JPA and squeezing-state techniques have rather small working bandwidths compared to that of cryogenics HEMT. It is also expected that the effective noise temperature can be further lowered by using JPA or TWPA as preamplifiers to protect the axion haloscope detector and beam splitter from thermal noise leaked from the high temperature stage, e.g., HEMT at $4\,$K ~\cite{Eichler2015}. It was proposed to detect the axion by a microwave single-photon detector~\cite{Pankratov2022}. We emphasize that it is also easy to combine the dual-path measurement setup with a single-photon detector, and it is expected that the SNR of axion detection reaches a limit from the temperature of the axion haloscope cavity itself. We think the best way would be to combine a dual-path scheme with broad-band, high-efficiency single-photon detectors for application in axion search in the future.

\section{Conclusions}
The axion cavity signal can be modeled quantum-mechanically because a single photon Fock state in the cavity is populated and then quickly decayed. We show that the single-photon $a\rightarrow \gamma$ transition rate is amplified by the cavity $Q$-factor, consistent with the classical picture. Since the axion cavity can be regarded as a quantum signal emitter, it is effective to employ the cross-power measurement in a dual-path interferometry setup instead of using a single channel receiver. Our dual-path HBT interferometry builds on this work's quantum level proof of the enhancement to exploit the single-photon cavity state, which differentiates the purpose from Ref.~\cite{McAllister:2015ofz}. The single-photon signal is splitted by a 50/50 beam splitter and then amplified and mixed by heterodyne detectors. With a digital analyzer, the measured cross-power effectively reduces the noise level due to the cancellation of uncorrelated noise in the two channels. Compared to the traditional single-channel readout, a high enhancement in the signal-to-noise ratio can be achieved with the current dual-path setup with the HEMT as the preamplifier. A higher enhancement for the JPA as the preamplifier is achievable because the amplified vacuum variance (the minimum is 0.5 photons) is avoided. Thus, combining the dual-path interferometry scheme with other techniques, e.g., single-photon detectors, could approach the noise temperature limited by the temperature of the axion haloscope cavity itself. The dual-path scheme can provide a substantially faster scanning rate for axion dark matter searches or, equivalently, a higher sensitivity of axion photon coupling. The second-order correlation function measurement can additionally provide a test of the single-photon nature of an axion-converted signal.
\section*{Data Availability}
The data used in this paper are available from authors on reasonable request.
\section*{Author Contributions}
These authors contributed equally: Q.Yang, Y.Gao, Z.Peng.
\section*{Competing Interests}
The authors declare no competing interests.
\section*{Acknowledgments}
The authors thank Pierre Sikvie, Giovanni Carugno, Lan Zhou, Yu-Xi Liu and Chang-Ling Zou for helpful discussions. Q.Y. is supported by NSFC under Grant No.~11875148 and No.~12150010. Y.G. is supported by NSFC under Grant No.~12150010 and is partially supported by the National R\&D Program of China, 2020YFC2201601. Z.P. is supported by NSFC under Grant Nos.~61833010,~12074117, and~12061131011. This research is supported in part by the Scientific Instrument Developing Project of the Chinese Academy of Sciences, Grant No. YJKYYQ20190049, and by the International Partnership Program of Chinese Academy of Sciences for Grand Challenges, Grant No. 112311KYSB20210012.

\bigskip

\bibliographystyle{ieeetr}
\renewcommand{\bibsection}{\textbf{References}}
\bibliography{refs}

\begin{thebibliography}{10}

\bibitem{Ade:2015xua}
P.~A.~R. Ade {\em et~al.}, ``{Planck 2015 results. XIII. Cosmological
  parameters},'' {\em Astron. Astrophys.}, vol.~594, p.~A13, 2016.

\bibitem{Peccei:1977hh}
R.~D. Peccei and H.~R. Quinn, ``{CP Conservation in the Presence of
  Instantons},'' {\em Phys. Rev. Lett.}, vol.~38, pp.~1440--1443, 1977.
\newblock [,328(1977)].

\bibitem{Peccei:1977ur}
R.~D. Peccei and H.~R. Quinn, ``{Constraints Imposed by CP Conservation in the
  Presence of Instantons},'' {\em Phys. Rev.}, vol.~D16, pp.~1791--1797, 1977.

\bibitem{Weinberg:1977ma}
S.~Weinberg, ``{A New Light Boson?},'' {\em Phys. Rev. Lett.}, vol.~40,
  pp.~223--226, 1978.

\bibitem{Wilczek:1977pj}
F.~Wilczek, ``{Problem of Strong $P$ and $T$ Invariance in the Presence of
  Instantons},'' {\em Phys. Rev. Lett.}, vol.~40, pp.~279--282, 1978.

\bibitem{Kim:1979if}
J.~E. Kim, ``{Weak Interaction Singlet and Strong CP Invariance},'' {\em Phys.
  Rev. Lett.}, vol.~43, p.~103, 1979.

\bibitem{Shifman:1979if}
M.~A. Shifman, A.~I. Vainshtein, and V.~I. Zakharov, ``{Can Confinement Ensure
  Natural CP Invariance of Strong Interactions?},'' {\em Nucl. Phys.},
  vol.~B166, pp.~493--506, 1980.

\bibitem{Zhitnitsky:1980tq}
A.~R. Zhitnitsky, ``{On Possible Suppression of the Axion Hadron Interactions.
  (In Russian)},'' {\em Sov. J. Nucl. Phys.}, vol.~31, p.~260, 1980.
\newblock [Yad. Fiz.31,497(1980)].

\bibitem{Dine:1981rt}
M.~Dine, W.~Fischler, and M.~Srednicki, ``{A Simple Solution to the Strong CP
  Problem with a Harmless Axion},'' {\em Phys. Lett.}, vol.~104B, pp.~199--202,
  1981.

\bibitem{Preskill:1982cy}
J.~Preskill, M.~B. Wise, and F.~Wilczek, ``{Cosmology of the Invisible
  Axion},'' {\em Phys. Lett.}, vol.~B120, pp.~127--132, 1983.
\newblock [,URL(1982)].

\bibitem{Abbott:1982af}
L.~F. Abbott and P.~Sikivie, ``{A Cosmological Bound on the Invisible Axion},''
  {\em Phys. Lett.}, vol.~B120, pp.~133--136, 1983.
\newblock [,URL(1982)].

\bibitem{Dine:1982ah}
M.~Dine and W.~Fischler, ``{The Not So Harmless Axion},'' {\em Phys. Lett.},
  vol.~B120, pp.~137--141, 1983.
\newblock [,URL(1982)].

\bibitem{Sikivie:1982qv}
P.~Sikivie, ``{Of Axions, Domain Walls and the Early Universe},'' {\em Phys.
  Rev. Lett.}, vol.~48, pp.~1156--1159, 1982.

\bibitem{Wilczek:1991jgb}
F.~Wilczek, ``{The Birth of Axions},'' 1991.

\bibitem{Crewther:1979pi}
R.~J. Crewther, P.~Di~Vecchia, G.~Veneziano, and E.~Witten, ``{Chiral Estimate
  of the Electric Dipole Moment of the Neutron in Quantum Chromodynamics},''
  {\em Phys. Lett. B}, vol.~88, p.~123, 1979.
\newblock [Erratum: Phys.Lett.B 91, 487 (1980)].

\bibitem{Baker:2006ts}
C.~A. Baker {\em et~al.}, ``{An Improved experimental limit on the electric
  dipole moment of the neutron},'' {\em Phys. Rev. Lett.}, vol.~97, p.~131801,
  2006.

\bibitem{Peccei:2006as}
R.~D. Peccei, ``{The Strong CP problem and axions},'' {\em Lect. Notes Phys.},
  vol.~741, pp.~3--17, 2008.

\bibitem{Kim:2008hd}
J.~E. Kim and G.~Carosi, ``{Axions and the Strong CP Problem},'' {\em Rev. Mod.
  Phys.}, vol.~82, pp.~557--602, 2010.
\newblock [Erratum: Rev.Mod.Phys. 91, 049902 (2019)].

\bibitem{Callan:1977gz}
C.~G. Callan, Jr., R.~F. Dashen, and D.~J. Gross, ``{Toward a Theory of the
  Strong Interactions},'' {\em Phys. Rev. D}, vol.~17, p.~2717, 1978.

\bibitem{Vafa:1984xg}
C.~Vafa and E.~Witten, ``{Parity Conservation in QCD},'' {\em Phys. Rev.
  Lett.}, vol.~53, p.~535, 1984.

\bibitem{Sikivie:2006ni}
P.~Sikivie, ``{Axion Cosmology},'' {\em Lect. Notes Phys.}, vol.~741,
  pp.~19--50, 2008.

\bibitem{Hertzberg:2008wr}
M.~P. Hertzberg, M.~Tegmark, and F.~Wilczek, ``{Axion Cosmology and the Energy
  Scale of Inflation},'' {\em Phys. Rev. D}, vol.~78, p.~083507, 2008.

\bibitem{DePanfilis:1987dk}
S.~De~Panfilis, A.~C. Melissinos, B.~E. Moskowitz, J.~T. Rogers, Y.~K.
  Semertzidis, W.~Wuensch, H.~J. Halama, A.~G. Prodell, W.~B. Fowler, and F.~A.
  Nezrick, ``{Limits on the Abundance and Coupling of Cosmic Axions at
  4.5-Microev \ensuremath{<} m(a) \ensuremath{<} 5.0-Microev},'' {\em Phys.
  Rev. Lett.}, vol.~59, p.~839, 1987.

\bibitem{Hagmann:1990tj}
C.~Hagmann, P.~Sikivie, N.~S. Sullivan, and D.~B. Tanner, ``{Results from a
  search for cosmic axions},'' {\em Phys. Rev. D}, vol.~42, pp.~1297--1300,
  1990.

\bibitem{Kahn:2016aff}
Y.~Kahn, B.~R. Safdi, and J.~Thaler, ``{Broadband and Resonant Approaches to
  Axion Dark Matter Detection},'' {\em Phys. Rev. Lett.}, vol.~117, no.~14,
  p.~141801, 2016.

\bibitem{Caldwell:2016dcw}
A.~Caldwell, G.~Dvali, B.~Majorovits, A.~Millar, G.~Raffelt, J.~Redondo,
  O.~Reimann, F.~Simon, and F.~Steffen, ``{Dielectric Haloscopes: A New Way to
  Detect Axion Dark Matter},'' {\em Phys. Rev. Lett.}, vol.~118, no.~9,
  p.~091801, 2017.

\bibitem{HAYSTAC:2018rwy}
L.~Zhong {\em et~al.}, ``{Results from phase 1 of the HAYSTAC microwave cavity
  axion experiment},'' {\em Phys. Rev. D}, vol.~97, no.~9, p.~092001, 2018.

\bibitem{Marsh:2018dlj}
D.~J.~E. Marsh, K.-C. Fong, E.~W. Lentz, L.~Smejkal, and M.~N. Ali, ``{Proposal
  to Detect Dark Matter using Axionic Topological Antiferromagnets},'' {\em
  Phys. Rev. Lett.}, vol.~123, no.~12, p.~121601, 2019.

\bibitem{Schutte-Engel:2021bqm}
J.~Sch\"utte-Engel, D.~J.~E. Marsh, A.~J. Millar, A.~Sekine, F.~Chadha-Day,
  S.~Hoof, M.~N. Ali, K.-C. Fong, E.~Hardy, and L.~\v{S}mejkal, ``{Axion
  quasiparticles for axion dark matter detection},'' {\em JCAP}, vol.~08,
  p.~066, 2021.

\bibitem{ADMX:2021nhd}
C.~Bartram {\em et~al.}, ``{Search for Invisible Axion Dark Matter in the
  3.3\textendash{}4.2\,\,\ensuremath{\mu}eV Mass Range},'' {\em Phys. Rev.
  Lett.}, vol.~127, no.~26, p.~261803, 2021.

\bibitem{Wilczek:1987mv}
F.~Wilczek, ``{Two Applications of Axion Electrodynamics},'' {\em Phys. Rev.
  Lett.}, vol.~58, p.~1799, 1987.

\bibitem{Essin:2008rq}
A.~M. Essin, J.~E. Moore, and D.~Vanderbilt, ``{Magnetoelectric polarizability
  and axion electrodynamics in crystalline insulators},'' {\em Phys. Rev.
  Lett.}, vol.~102, p.~146805, 2009.

\bibitem{Li:2009tca}
R.~Li, J.~Wang, X.~Qi, and S.-C. Zhang, ``{Dynamical Axion Field in Topological
  Magnetic Insulators},'' {\em Nature Phys.}, vol.~6, p.~284, 2010.

\bibitem{Xiao:2017mol}
D.~Xiao {\em et~al.}, ``{Realization of the Axion Insulator State in Quantum
  Anomalous Hall Sandwich Heterostructures},'' {\em Phys. Rev. Lett.},
  vol.~120, no.~5, p.~056801, 2018.

\bibitem{PhysRevLett.105.190404}
A.~Bermudez, L.~Mazza, M.~Rizzi, N.~Goldman, M.~Lewenstein, and M.~A.
  Martin-Delgado, ``Wilson fermions and axion electrodynamics in optical
  lattices,'' {\em Phys. Rev. Lett.}, vol.~105, p.~190404, Nov 2010.

\bibitem{Nenno:2020ujq}
D.~M. Nenno, C.~A.~C. Garcia, J.~Gooth, C.~Felser, and P.~Narang, ``{Axion
  physics in condensed-matter systems},'' {\em Nature Rev. Phys.}, vol.~2,
  no.~12, pp.~682--696, 2020.

\bibitem{Sikivie:1983ip}
P.~Sikivie, ``{Experimental Tests of the Invisible Axion},'' {\em Phys. Rev.
  Lett.}, vol.~51, pp.~1415--1417, 1983.
\newblock [Erratum: Phys.Rev.Lett. 52, 695 (1984)].

\bibitem{Brown:1956zza}
R.~H. Brown and R.~Q. Twiss, ``{Correlation between Photons in two Coherent
  Beams of Light},'' {\em Nature}, vol.~177, pp.~27--29, 1956.

\bibitem{JPA2011}
M.~Hatridge, R.~Vijay, D.~H. Slichter, J.~Clarke, and I.~Siddiqi, ``Dispersive
  magnetometry with a quantum limited squid parametric amplifier,'' {\em
  Physical Review B}, vol.~83, Apr 2011.

\bibitem{Menzel2010}
E.~P. Menzel, F.~Deppe, M.~Mariantoni, M.~. Araque~Caballero, A.~Baust,
  T.~Niemczyk, E.~Hoffmann, A.~Marx, E.~Solano, and R.~Gross, ``Dual-path state
  reconstruction scheme for propagating quantum microwaves and detector noise
  tomography,'' {\em Physical Review Letters}, vol.~105, Aug 2010.

\bibitem{Bozyigit2010}
D.~Bozyigit, C.~Lang, L.~Steffen, J.~Fink, C.~Eichler, M.~Baur, R.~Bianchetti,
  P.~Leek, S.~Filipp, M.~da~Silva, A.~Blais, and A.~Wallraff, ``{Antibunching
  of microwave-frequency photons observed in correlation measurements using
  linear detectors},'' {\em Nature Phys}, vol.~7, pp.~154--158, 2011.

\bibitem{Peng2016}
Z.~H. Peng, S.~E. de~Graaf, J.~S. Tsai, and O.~V. Astafiev, ``Tuneable
  on-demand single-photon source in the microwave range,'' {\em Nature
  Communications}, vol.~7, Aug 2016.

\bibitem{Zhou2020}
Y.~Zhou, Z.~Peng, Y.~Horiuchi, O.~Astafiev, and J.~Tsai, ``Tunable microwave
  single-photon source based on transmon qubit with high efficiency,'' {\em
  Physical Review Applied}, vol.~13, Mar 2020.

\bibitem{Sikivie:1985yu}
P.~Sikivie, ``{Detection Rates for 'Invisible' Axion Searches},'' {\em Phys.
  Rev. D}, vol.~32, p.~2988, 1985.
\newblock [Erratum: Phys.Rev.D 36, 974 (1987)].

\bibitem{Beutter:2018xfx}
M.~Beutter, A.~Pargner, T.~Schwetz, and E.~Todarello, ``{Axion-electrodynamics:
  a quantum field calculation},'' {\em JCAP}, vol.~02, p.~026, 2019.

\bibitem{Sikivie:2001fg}
P.~Sikivie, ``{Evidence for ring caustics in the Milky Way},'' {\em Phys. Lett.
  B}, vol.~567, pp.~1--8, 2003.

\bibitem{Armendariz-Picon:2013jej}
C.~Armendariz-Picon and J.~T. Neelakanta, ``{How Cold is Cold Dark Matter?},''
  {\em JCAP}, vol.~03, p.~049, 2014.

\bibitem{Fitzpatrick2015}
R.~Fitzpatrick, ``{Quantum Mechanics},'' {\em World Scientific}, p.~181, 2015.

\bibitem{Foster:2017hbq}
J.~W. Foster, N.~L. Rodd, and B.~R. Safdi, ``{Revealing the Dark Matter Halo
  with Axion Direct Detection},'' {\em Phys. Rev. D}, vol.~97, no.~12,
  p.~123006, 2018.

\bibitem{Qian:2021}
Z.~Qian {\em et~al.}, ``{Spontaneous emission in micro- or nanophotonic
  structures},'' {\em PhotoniX}, vol.~2, p.~21, 2021.

\bibitem{PhysRevX.10.021038}
R.~Lescanne, S.~Del\'eglise, E.~Albertinale, U.~R\'eglade, T.~Capelle,
  E.~Ivanov, T.~Jacqmin, Z.~Leghtas, and E.~Flurin, ``Irreversible qubit-photon
  coupling for the detection of itinerant microwave photons,'' {\em Phys. Rev.
  X}, vol.~10, p.~021038, May 2020.

\bibitem{Blais:2020wjs}
A.~Blais, A.~L. Grimsmo, S.~M. Girvin, and A.~Wallraff, ``{Circuit quantum
  electrodynamics},'' {\em Rev. Mod. Phys.}, vol.~93, no.~2, p.~025005, 2021.

\bibitem{PhysRevLett.126.141302}
A.~V. Dixit, S.~Chakram, K.~He, A.~Agrawal, R.~K. Naik, D.~I. Schuster, and
  A.~Chou, ``Searching for dark matter with a superconducting qubit,'' {\em
  Phys. Rev. Lett.}, vol.~126, p.~141302, Apr 2021.

\bibitem{Bartram:2021ysp}
C.~Bartram {\em et~al.}, ``{Dark matter axion search using a Josephson
  Traveling wave parametric amplifier},'' {\em Rev. Sci. Instrum.}, vol.~94,
  no.~4, p.~044703, 2023.

\bibitem{Alesini:2020vny}
D.~Alesini {\em et~al.}, ``{Search for invisible axion dark matter of mass
  m$_a=43~\mu$eV with the QUAX--$a\gamma$ experiment},'' {\em Phys. Rev. D},
  vol.~103, no.~10, p.~102004, 2021.

\bibitem{Caves1982}
C.~M. Caves, ``Quantum limits on noise in linear amplifiers,'' {\em Phys. Rev.
  D}, vol.~26, pp.~1817--1839, Oct 1982.

\bibitem{Macklin2015}
C.~Macklin, K.~O’Brien, D.~Hover, M.~E. Schwartz, V.~Bolkhovsky, X.~Zhang,
  W.~D. Oliver, and I.~Siddiqi, ``A near–quantum-limited josephson
  traveling-wave parametric amplifier,'' {\em Science}, vol.~350, pp.~307--310,
  Oct. 2015.

\bibitem{Eichler2015}
C.~Eichler, J.~Mlynek, J.~Butscher, P.~Kurpiers, K.~Hammerer, T.~Osborne, and
  A.~Wallraff, ``Exploring interacting quantum many-body systems by
  experimentally creating continuous matrix product states in superconducting
  circuits,'' {\em Physical Review X}, vol.~5, Dec 2015.

\bibitem{Silva2010}
M.~P. da~Silva, D.~Bozyigit, A.~Wallraff, and A.~Blais, ``Schemes for the
  observation of photon correlation functions in circuit qed with linear
  detectors,'' {\em Phys. Rev. A}, vol.~82, pp.~043804--, Oct. 2010.

\bibitem{Backes2021}
K.~M. Backes, D.~A. Palken, S.~A. Kenany, B.~M. Brubaker, S.~B. Cahn,
  A.~Droster, G.~C. Hilton, S.~Ghosh, H.~Jackson, S.~K. Lamoreaux, and et~al.,
  ``A quantum enhanced search for dark matter axions,'' {\em Nature}, vol.~590,
  p.~238–242, Feb 2021.

\bibitem{Pankratov2022}
A.~Pankratov, L.~Revin, A.~Gordeeva, A.~Yablokov, L.~Kuzmin, and E.~Il’ichev,
  ``Towards a microwave single-photon counter for searching axions,'' {\em npj
  Quantum Inf}, vol.~8, p.~61, 2022.

\bibitem{McAllister:2015ofz}
B.~T. McAllister, S.~R. Parker, E.~N. Ivanov, and M.~E. Tobar,
  ``{Cross-Correlation Signal Processing for Axion and WISP Dark Matter
  Searches},'' {\em IEEE Trans. Ultrason. Ferroelectr. Freq. Control}, vol.~66,
  no.~1, pp.~236--243, 2019.

\end{thebibliography}

\end{document}